\documentclass{mn2e}
\usepackage{graphicx}
\newcommand{\text}[1]{\quad\mbox{#1}\quad}

\newcommand{\sub}[1]{_{\mbox{\tiny #1}}}
\title{Magnetar-energized supernova explosions and GRB-jets}
\author[S.~S. Komissarov \& M.V.Barkov]{Serguei
S.~Komissarov,$^{1}$\thanks{E-Mail:~serguei@maths.leeds.ac.uk~(SSK);
bmv@maths.leeds.ac.uk (MVB)}
Maxim V.~Barkov,$^{1,2}$\footnotemark[1]\\
$^{1}$Department of Applied Mathematics, The University of Leeds,
Leeds, LS2 9GT\\
$^{2}$Space Research Institute, 84/32 Profsoyuznaya Street, Moscow
117997, Russia}
\begin{document}
\date{Received/Accepted}
\maketitle

\begin{abstract}
In this paper we report on the early evolution of a core-collapse supernova
explosion following the birth of a magnetar with the dipolar magnetic field of
$B=10^{15}$G and the rotational period of $2$ms, which was studied by means of
axisymmetric general relativistic MHD simulations. In this study we use
realistic EOS and take into account the cooling and heating associated with
emission, absorption, annihilation, and scattering of neutrinos, the neutrino
transport is treated in the optically-thin regime.  The supernova explosion is
initiated via introducing into the initial solution the ``radiation bubble'',
whose total thermal energy is comparable with the typical energy of supernova
ejecta.  The numerical models exhibit highly collimated magnetically-driven jets
very early on. The jets are super-Alfv\'enic but remain sub-fast until the end
of the simulations (t=0.2s). The power released in the jets is about
$3\times10^{50}$erg/s which implies the spin-down time of $\simeq 37$s. The
total rotational energy of the magnetar, $E\simeq 10^{52}$erg, is sufficient to
drive a hypernova but it is not clear as to how large a fraction of this energy
can be transfered to the stellar ejecta. Given the observed propagation speed
of the jets, $v_p\simeq 0.17c$, they are expected to traverse the progenitor in
few seconds and after this most of the released rotational energy would be
simply carried away by these jets into the surrounding space.  3-dimensional
effects such as the kink mode instability may reduce the jet propagation speed
and increase the amount of energy transferred by the jets to the supernova
ejecta.  Our results provide the first more or less self-consistent numerical
model of a central engine capable of producing, in the supernova setting and on
a long-term basis, collimated jets with sufficient power to explain long
duration GRBs and their afterglows. Although the flow speed of our jets is
relatively low, only $v_j\simeq 0.5c$, the cooling of proto-neutron star will
eventually result in much higher magnetization of its magnetosphere and
ultra-relativistic asymptotic speeds of the jets. Given the relatively long
cooling time-scale we still expect the jets to be only weakly relativistic by
the time of break out.  This leads to a model of GRB jets with systematic
longitudinal variation of Lorentz factor which may have specific observational
signatures both in the prompt and the afterglow emission.  The simulations also
reveal quasi-periodic ejection of plasma clouds into the jet on a time-scale of
20ms related to the large-scale global oscillation of magnetar's magnetosphere
caused by the opening-closing of the dead zone field lines. These kind of
central engine variability may be partly responsible for the internal shocks of
GRB jets and the short-time variability of their gamma-ray emission.           
\end{abstract}
                                                                                          
\begin{keywords}
supernovae: general -- stars: neutron -- gamma-rays: bursts  -- methods: numerical -- MHD --
relativity
\end{keywords}
                                                                                          
\section{Introduction}
\label{introduction}

After decades of intensive research the exact mechanism of core-collapse
supernovae (SNe) still remains a mystery. It is widely believed that the
explosion is driven by neutrinos emitted by the proto-neutron star (PNS) formed
as the result of the collapse \cite{bethe}.  However, the attempts to reproduce
core-collapse SNe in computer simulations have encountered severe problems. The
reason for the failures becomes more or less clear when one compares the total
energy radiated in the form of neutrinos, $\simeq10^{53}$erg, with the typical
energy of supernovae, $\simeq10^{51}$erg.  This tells us that in order to obtain
a reliable answer the computational error in the energy budget has to be below
$1\%$, which a is very demanding condition.  The neutrino transport, of six
different species, has to be treated with great accuracy including the processes
effecting heating and cooling rates.

The alternative magnetic mechanism of core-collapse SNe has been around for a
while, first proposed by Bisnovatyi-Kogan\shortcite{BK70} and LeBlanc \&
Wilson\shortcite{LW70}, whose numerical simulations were miles ahead of their
time. For three decades this mechanism was not taken seriously and only
occasionally efforts were made to develop it further \cite{BPS76,M76,S84}.  Now
it is experiencing ``renaissance''
\cite{WIHW00,WMW02,AWML03,YS04,KSYS04,MYKS04,TKNS04,ABM05,AW5,Pr05,MBA06,OAM06,SLSS06,MSS07,N07,BDLOM07},
thanks to the slow and difficult progress of the neutrino models of SNe, the
development of robust numerical methods for MHD, the accumulated evidence for
asphericity of supernovae \cite{Kho99,W02,W03} \footnote{ 
Based on the results of 2D axisymmetric numerical simulations, it has been proposed that 
strongly aspherical explosions may arise in non-magnetic models too \cite{BMD03,BL06}. 
3D simulations are needed to test this idea.   
}, the high collimation of flows
associated with GRBs (e.g. Piran 2005a,b) and the increasing popularity of the
magnetic mechanism for the origin of other astrophysical jets. The results show
that magnetic field can facilitate explosions of standard power and even more
powerful explosions provided the magnetic field is sufficiently strong.

The energy of high-velocity supernovae or hypernovae (HNe) exceeds that of
normal core-collapse supernovae by about ten times (e.g. Nomoto et al. 2004).
The few SNe that have been reliably identified with GRBs are all HNe.  Although,
the statistics is very poor and is subject to observational bias towards most
powerful and hence brightest events, the connection between HNe and GRBs is
widely accepted. The physical mechanism of HNe is not established and there is
no shortage of competing theoretical models.  The most popular one, the
collapsar model, relates HNe with collapse of massive rapidly rotating stars
\cite{MW99}. In this model, the stellar core collapses straight into a black
hole (which used to be considered as an indication of ``failed'' supernova
explosion) but the stellar envelope does not and forms a hyper-accreting disk
around the hole. The energy released in the disk can be very large, sufficient
to drive HN and GRB, but the way it is transferred to the supernova ejecta and
the GRB jets remains to be determined. It is widely assumed that GRB jets are
powered by the energy released in the funnel region of the disk via annihilation
of neutrinos \cite{RJ99,NPK01}. However, taking into account opacity in the inner regions of the 
disk significantly reduces the efficiency of this mechanism \cite{DPN02,N07}. 
Alternatively, as it has been suggested by many authors, the GRB-jets in the collapsar 
scenarion can be powered via magnetic mechanism tapping the rotational energy of 
the black hole and/or the accretion disk (e.g. Lyutikov \& Blandford~2003, 
Vlahakis \& Konigl~2003, Proga et al.,~2003).

The energy of HNe ejecta is of the same magnitude as the rotational energy of a
neutron star (NS) with very short period of rotation, 1-2 milliseconds. This
energy can be released via magnetic braking within a very short period of time,
from few seconds to few hundreds of a second, provided the magnetic field is very
strong, of order $B_{\mbox{\tiny NS}} \simeq 10^{15}$G \cite{T06,B06}. This
value is three orders of magnitude above the typical strength of magnetic field
of normal neutron stars and would be considered a huge over-stretch not so long
time ago. However, the identification of Soft Gamma Ray Repeaters and X-ray
Anomalous Pulsars with magnetars, neutron stars with magnetic field
$B_{\mbox{\tiny NS}}\simeq 10^{14}-10^{15}$G \cite{TD95,WT04}, shows that the
magnetar model may be a viable alternative to the collapsar model of HNe as well
as GRBs \cite{U92,TCQ04,T06,MTQ07,UM06}.

The origin of magnetar's magnetic field is still debated. One possibility is the
generation of unusually strong magnetic field in convective cores of some
pre-supernova stars. During the core collapse the poloidal field is amplified
like $\propto r^{-2}$ and in order to reach the required $B_{\mbox{\tiny
NS}}\simeq 10^{15}$G after the collapse the original field in the core, on the
scale of $~10^3$km, should be of order $B_{\mbox{\tiny C}} \simeq 10^{11}$G. The
collapse of magnetized rotating stellar cores has been studied by many groups.
The general conclusion from the latest parameter studies is that only for
$B_{\mbox{\tiny C}} \ge 10^{11}$G the magnetic field is strong enough to
influence the dynamics of supernovae explosions \cite{OAM06,BDLOM07}. The key
feature of these explosions is strong differential rotation and generation of
toroidal magnetic field whose pressure actually drives the explosion. The hoop
stress of this field also ensures anisotropy of the explosion - it is more
powerful along the rotational axis.
 
On the other hand, the super-strong magnetic field of magnetars could be
generated via $\alpha$-$\Omega$-dynamo in the convective PNS
\cite{DT92,TD93}. In this theory the strength of saturated field strongly
depends on the rotational period - shorter period leading to stronger field.  In
order to generate the dipolar field of $B_{\mbox{\tiny NS}}\simeq 10^{15}$G the
period of PNS should be very short, not much higher than few milliseconds
\cite{DT92}. Thus, this theory unites both basic conditions required to produce
a hypernova in the magnetar model, rapid rotation and super-strong magnetic
field, in one.  On the downside, the PNS is no longer magnetically coupled to
the envelope and the transfer of its rotational energy to the stellar envelope
is more problematic.  
Turbulence required for the magnetic dynamo action can also be generated via
the magneto-rotational instability (MRI, Balbus \& Hawley,1991). Calculations based 
on the linear theory show that in the supernova context strong saturation field can be 
reached very quickly on the time scale of only several tens of rotational periods 
\cite{AWML03}.

Thompson et al.\shortcite{TCQ04} proposed a model of magnetically driven HN
explosions and GRB where the usual delayed neutrino-driven mechanism plays the
role of bomb detonator.  The neutrino-driven blast creates a rarefaction around
the newly-born magnetar, its magnetosphere expands and eventually develops a
magnetically driven pulsar wind (PW).  This wind acts as a piston on the
expanding supernova envelope, that is the layer of compressed stellar material behind 
the supernova shock,  and energizes it beyond the level of normal SN.

Metzger et al.\shortcite{MTQ07} generalized the 1D model of equatorial
magnetically-driven wind by Weber \& Davis\shortcite{WD67} to fit the conditions
of PNS.  Axisymmetric magnetized winds of magnetars with various degree of mass
loading have been studied in Bucciantini et al.\shortcite{B06} via 2D numerical
simulations. They concluded that during the initial phase (Kelvin-Helmholtz
timescale) when the wind is already magnetically-driven but still
non-relativistic the rotational energy loss rate is $L\simeq 4\times10^{51}
B_{15}^2 P_1^{-5/3}$erg/s, where $B_{15}$ is the magnetic field of the PNS in
$10^{15}$G and $P_1$ is its period in milliseconds.  Bucciantini et
al.\shortcite{B07a} applied the model of pulsar wind nebula (PWN) by Begelman \&
Li~\shortcite{BL92} to study the late phases of hypernova explosion, $t>1$s.
They used the thin-shell approximation to describe the dynamics of the supernova
shock driven by the anisotropic pressure of the magnetized bubble, ``baby pulsar 
wind nebula'', created inside
the exploding star by the magnetar wind. They concluded that when the magnetic energy
in the bubble exceeds $\simeq 20\%$ of the thermal energy the shell expansion
proceeds in a highly anisotropic fashion and eventually an axial channel is
created inside the stellar envelope. They proposed that later this channel
collimates the ultra-relativistic wind from cooled magnetars into a pair of GRB jets.
In the follow up paper \cite{B07b} they studied the interaction of the 
super-fast magnetar wind with the supernova envelope via fully relativistic 
MHD simulations and arrived to similar conclusions.    
For the wind model to become applicable the size of the ``central cavity'' must
exceed, and likely many times, that of the fast surface as found in the
unconfined case. Prior to this one would expect a somewhat different
dynamics. 

The most pessimistic possibility is a rigidly rotating magnetosphere
in which case no extraction of rotational energy takes place.  However, the
theory of relativity places a upper limit on the size of such a
magnetosphere. Indeed, the magnetosphere whose size exceeds the light cylinder
(LC) radius cannot be rigidly rotating as this would imply superluminal motion
of magnetospheric plasma and should become differentially rotating.  This limit
is particularly relevant for the magnetospheres of millisecond pulsars as the
radius of LC, $R\sub{LC}\simeq 50 P_1$~km, is only few times the radius of the
star itself. Once the differential rotation sets up it amplifies the toroidal
field, the magnetic pressure grows and promotes further expansion of the
magnetosphere. Uzdensky \& MacFadyen\shortcite{UM06} suggested that this factor
along can explain supernova explosions but they did not elaborate ways to
achieve the initial expansion beyond LC. Moreover, they assumed low mass loading
of magnetic field lines, more suitable for NS than hot proto-NS, and concluded
that the magnetospheric plasma will exhibit ultra-relativistic rotation in the
equatorial plane.  In any case, this argument suggests that the magnetic braking
should begin to operate at the very early stages of the ``detonated''
magnetically-driven explosions, well before the central cavity expands beyond
the fast surface of unconfined magnetar wind and this may to have important
implications for their later development.

In this paper we describe an attempt to explore the early phase ($t\le 1$s) of
such explosions via axisymmetric General Relativistic MHD simulations.  Our main
goals are to study the basic dynamics of this phase, to determine the spin-down
power, and to assess the implications for the later evolution.  We use an upwind
conservative scheme that based on a linear Riemann solver and uses the
constrained transport method to evolve the magnetic field.  The details of this
numerical method and various tests are described in
Komissarov\shortcite{K99,K04b,K06}.  In Sec.2 we describe the details of
numerical experiments, including initial and boundary conditions, Sec.3
describes the results of simulations, and in Sec.4 we discuss their implications
for the model of magnetar-driven supernovae explosions and GRB jets.
Our conclusions are listed in Sec.5.

\section{Simulation Setup}
\label{setup}

The ultimate hypernovae/supernovae simulations should trace all phases the
stellar explosions, the onset of core collapse, core bounce, delayed explosion,
generation of magnetic field, formation of magnetic cavity, and break out of
magnetically-driven jets.  Unfortunately this is still beyond our current
capabilities and we have to make some simplifications.  Here we start our
simulations from the point when strong magnetic field has just appeared from
surface of already formed PNS.  To account for the previous stages of the
explosion we assume that PNS is surrounded by a "radiation bubble" with a large
amount of energy deposited by neutrino in form of heat. The bubble is surrounded
by the collapsing stellar envelope.

The gravitational attraction of PNS is introduced via Schwarzschild metric in
Boyer-Lindquist coordinates, $\{\phi,r,\theta\}$.  The two-dimensional
computational domain is $(r_0<r<r_1)\times(0<\theta<\pi)$, where $r_0$
corresponds to the radius of PNS and $r_1=10^4$km. The total mass within the
domain is small compared to the mass of PNS that allows us to ignore its
self-gravity. The grid is uniform in $\theta$ where it has 200 cells and almost
uniform in $\log(r)$ where it has 430 cells, the linear cell size being the same
in both directions.

\subsection{PNS}

The evolution of a non-rotating proto-neutron star of mass $M=1.4M_{\sun}$ 
have been studied in details by Burrows \& Lattimer~\shortcite{BL86}. 
According to this study the deleptonization time-scale is rather long, it takes 
around 5 seconds before half the leptons inside $M(r)<0.5M$ are lost. However, the 
contraction time-scale is much shorter - it takes only $\simeq 250$ms for the star 
radius to decrease down to $\simeq 15$km (see fig.5 in Burrows \& Lattimer, 1986). 
Using these results we fix the mass and radius of PNS to $M=1.4M_{\sun}$ and
$r_0=15$km respectively, and we choose the rotation period to be $P=2$ms.  
To set the boundary conditions on the PNS
surface we utilize the results of high resolution 1D numerical models of PNS
winds \cite{TBM01,MTQ07}.  The gas temperature of ``ghost cells'' is to
$T_0=4$~MeV, which is typical for a hot newly-born PNS, and their density is
$\rho_0=3\times10^9\mbox{g}\mbox{cm}^{-3}$.  This density corresponds to the
interface between the very thin exponential atmosphere of PNS and its wind.  The
initial magnetic field is purely poloidal with the azimuthal component of vector potential
\begin{equation}
A_\phi= f(r)\frac{B_0r_0^3}{2r} \sin^2\theta,
\label{pot}
\end{equation}
where
\begin{equation}
f(r)= \left\{
   \begin{array}{rcl}
     1-((r-r_0)/(r_c-r_0))^2 &\mbox{if}& r<r_c\\ 0 &\mbox{if}& r>r_c,
   \end{array}
   \right.
\label{mask}
\end{equation} 
$B_0=10^{15}$G and $r_c=1.8r_0$\footnote{Here we give the component of vector potential 
in non-normalized coordinate basis. Everywhere else in the paper the components 
of vectors are measured using normalized bases.}. For $f(r)=1$ equation~\ref{pot} describes
magnetic dipole of strength $B_0$ at $r=r_0$, $\theta=0$. The masking
function~(\ref{mask}) ``moves'' the magnetic field lines inside the sphere of
radius $r_c$ without changing the magnetic flux distribution over the PNS
surface. This distribution is preserved for the whole duration of simulations.
The radial velocity of ghost cells is set to zero and the mass outflow through
the boundary is found via solving the Riemann problem at the boundary interface.

We assume that neutrino luminosity in each species is given by the black body
formula

\begin{equation}         
 L_\nu=1.20\times 10^{26} \pi r_0^2 c \left(\frac{T_0}{1\mbox{MeV}}\right)^4
\frac{\mbox{erg}}{\mbox{s}} \simeq 6.5\times 10^{51} \frac{\mbox{erg}}{\mbox{s}}
\label{lumin}
\end{equation}
and that the mean neutrino energy is $e_\nu=3.15\,T_0=12.6$~MeV. This
oversimplification is unlikely to have a strong effect on the largely
magnetically driven outflows from PNS.

\begin{figure*}
\includegraphics[width=85mm]{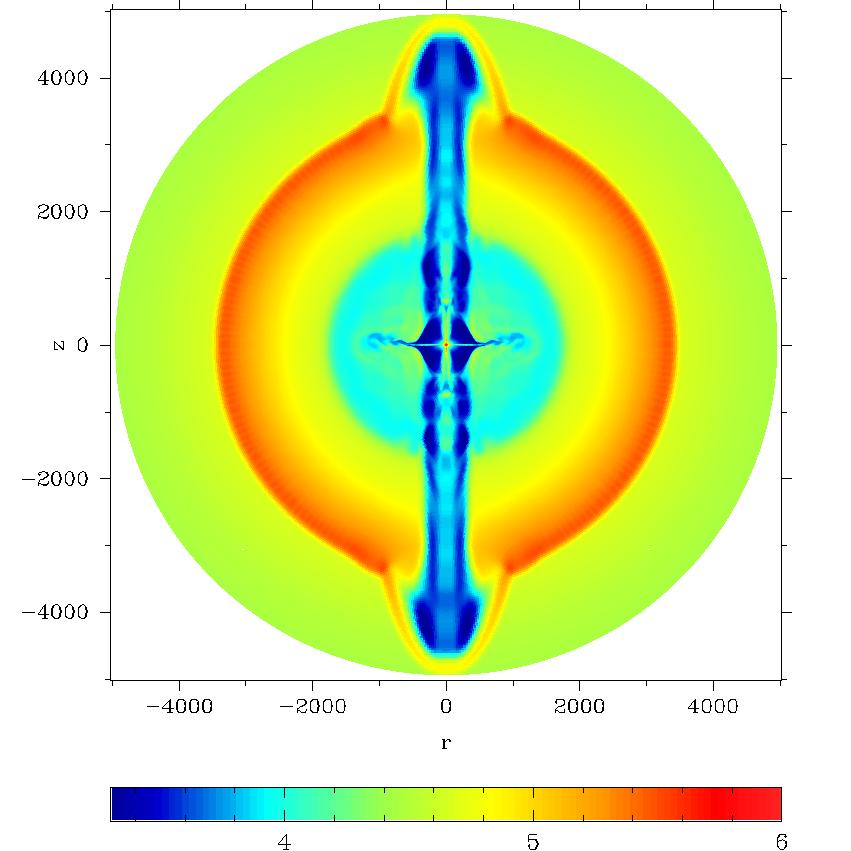}
\includegraphics[width=85mm]{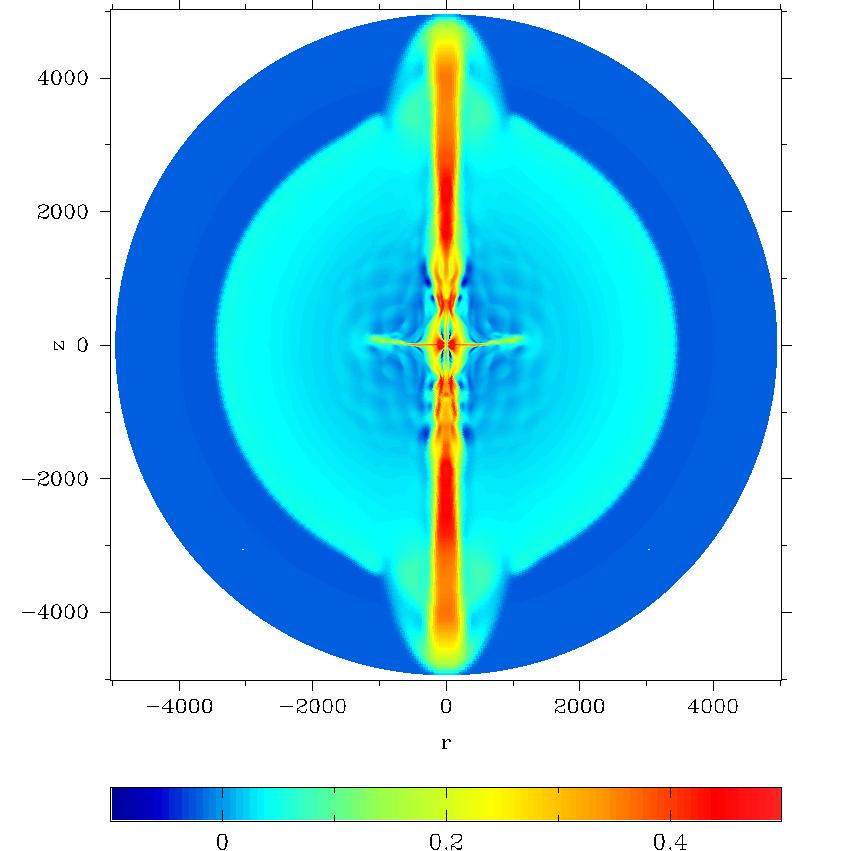}
\caption{Model A at time $t\simeq 200$~ms. {\it Left panel:} $log_{10}\rho$
measured in $\mbox{g~cm}^{-3}$.  {\it Right panel:} Radial velocity,
$v^{r}/c$. The unit length in all figures in this paper is ${\cal L}=GM/c^2 \simeq
2$km.  The dynamic range of colour plots does not always reflect the full range
of variation of the represented quantity but is rather selected to make more
revealing images. }
\label{f1}
\end{figure*}

\begin{figure*}
\includegraphics[width=85mm]{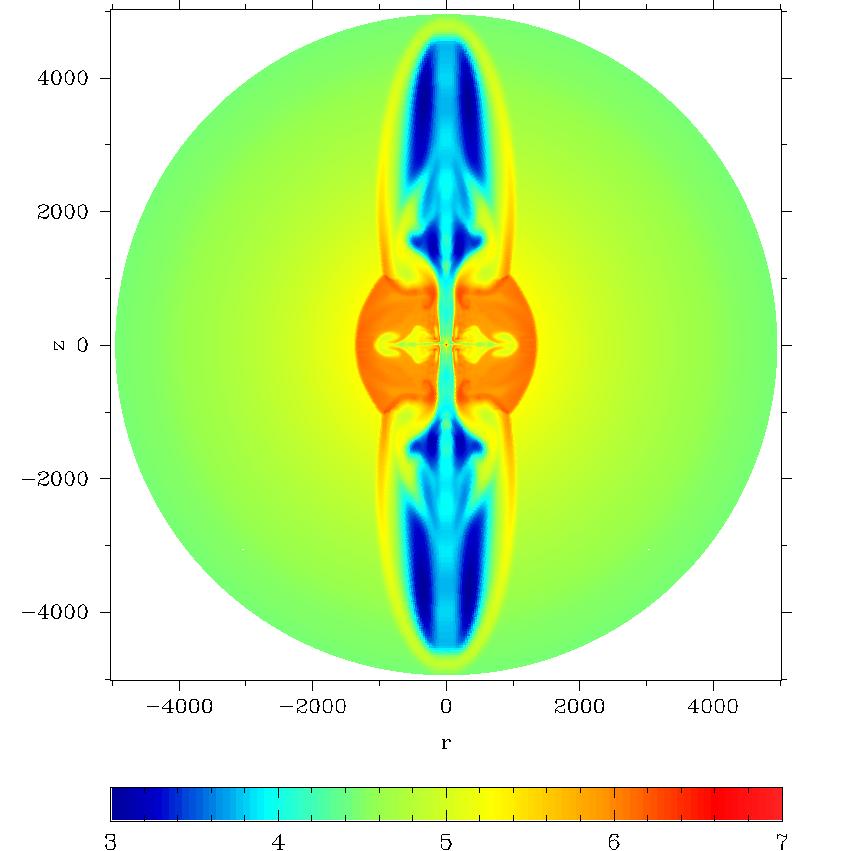}
\includegraphics[width=85mm]{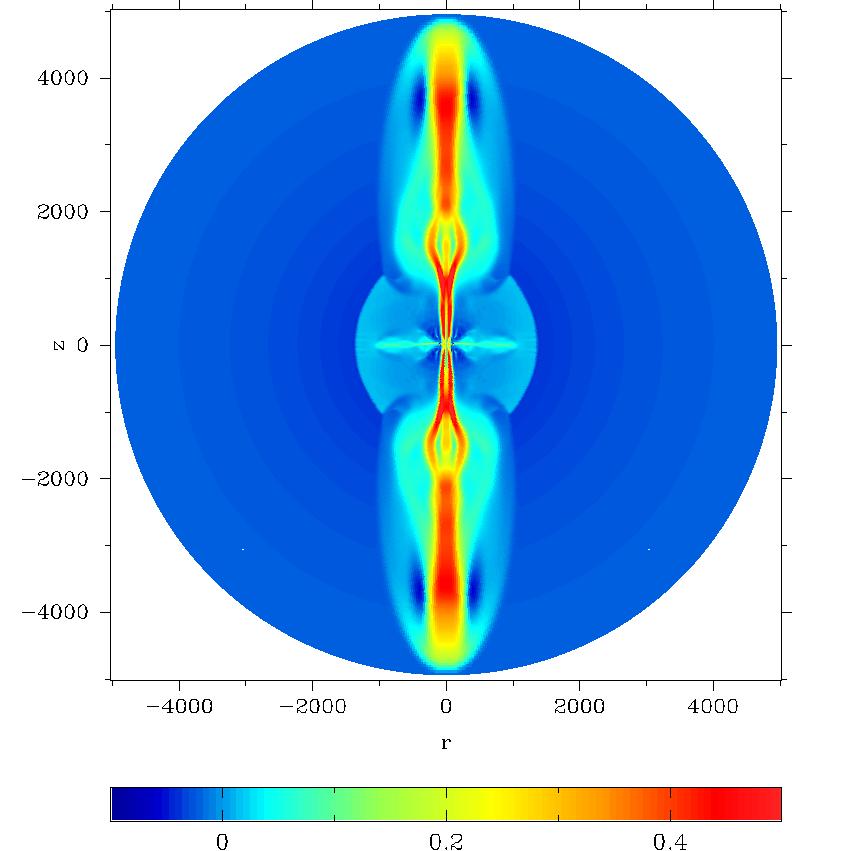}
\caption{Same as in figure~\ref{f1} but for model B at $t\simeq200$~ms.}
\label{f2}
\end{figure*}
\begin{figure*}
\includegraphics[width=85mm]{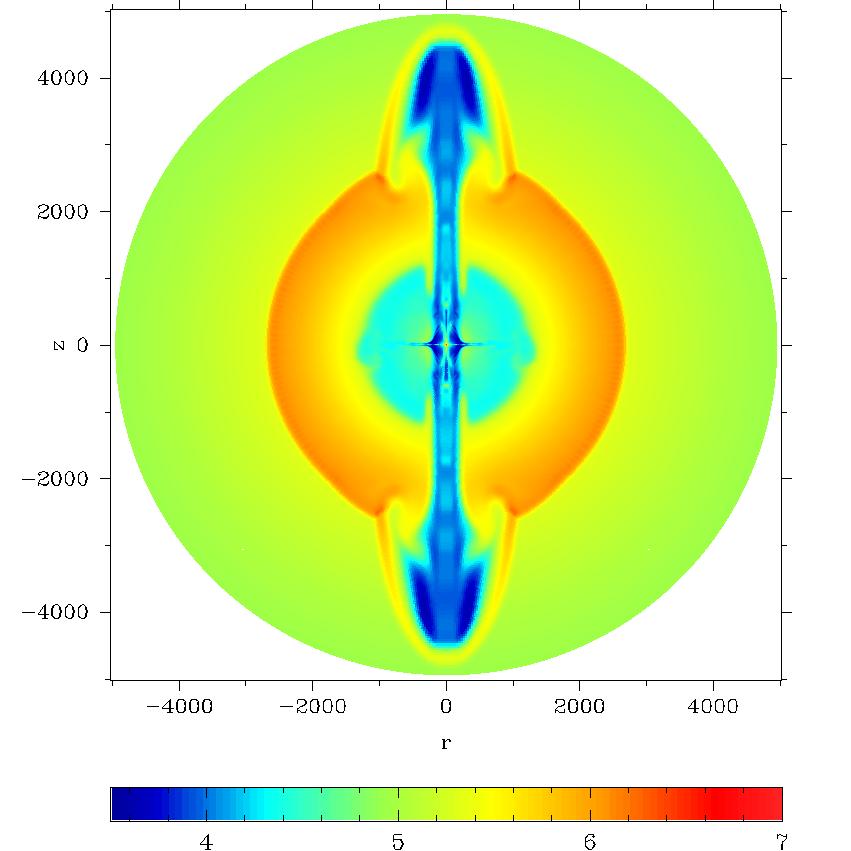}
\includegraphics[width=85mm]{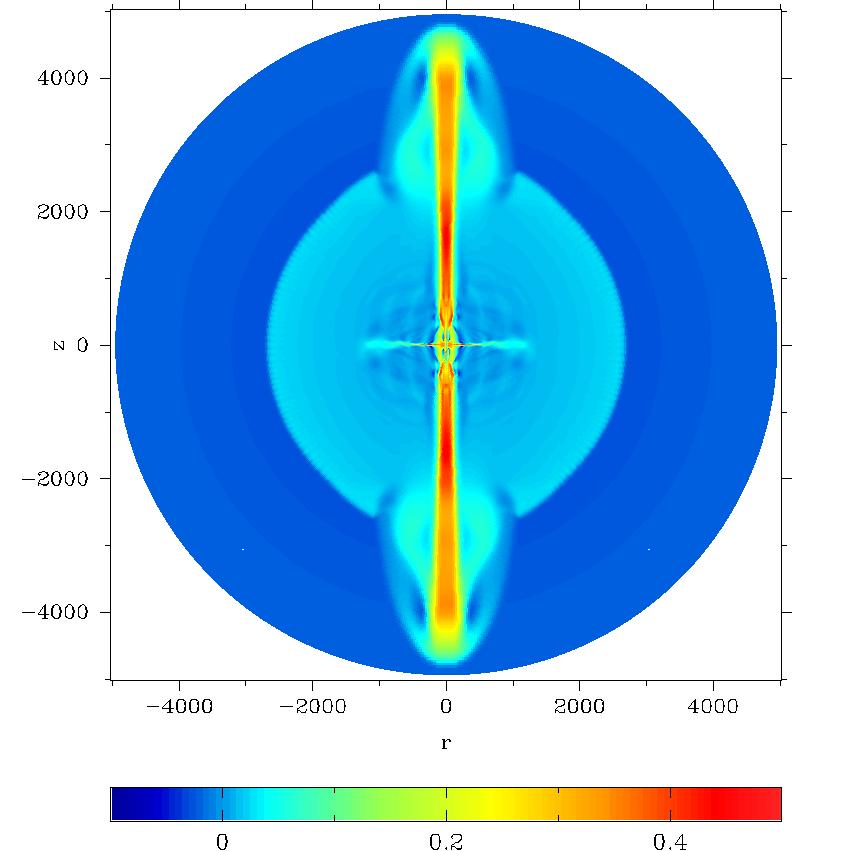}
\caption{Same as in figures~\ref{f1},\ref{f2} but for model C at $t\simeq277$~ms.}
\label{f2a}
\end{figure*}

\subsection{Collapsing star}

Here we adopt the simple free-fall model by Bethe~\shortcite{bethe}. According
to this model the radial velocity is
\begin{equation}
 v^r_{\mbox{ff}}=(2GM/r)^{1/2},
\label{b1}
\end{equation}
and the density is

\begin{equation} 
\rho = C_1\times10^7 \left(\frac{t_s}{1\mbox{~s}}\right)^{-1} \left(\frac{r}{10^7
        \mbox{cm}}\right)^{-3/2} \mbox{g~cm}^{-3},
\label{b2}
\end{equation}    
where $C_1$ is a coefficient between 1 and 10 \cite{bethe} 
and $t_s$ is the time since the onset of collapse.  
The corresponding accretion rate and ram pressure are

\begin{equation}
\dot{M} = 0.038~C_1 \left(\frac{M}{1.4M_{\sun}}\right)^{1/2}
\left(\frac{t_s}{1\mbox{~s}}\right)^{-1} M_{\sun} \mbox{s}^{-1},
\label{b3}
\end{equation}

\begin{equation}
p_{\mbox{\tiny ram}} = 3.7\times 10^{26} C_1 \frac{M}{1.4M_{\sun}}
\left(\frac{t_s}{1\mbox{~s}}\right)^{-1}
\left(\frac{r}{10^7\mbox{cm}}\right)^{-5/2} \frac{\mbox{g}}{\mbox{cm~s}^{2}}
\label{b4}
\end{equation}
respectively. 
Since in the simulations we fix the core mass to $M=1.4M_{\sun}$ the model is 
fully determined by the combination $C_1/t_s$.  
For a core of radius $r_c=10^9$cm and mass 1.4$N_{\sun}$ the collapse duration   
can be estimated as $t_c=2r_c/3v_{\mbox{ff}}\simeq 1$s. The time-scale of the 
delayed explosion is likely to be around few tens of a second after bounce \cite{bethe}. 
Thus, the value of $t_s$ is relatively well constrained and we fixed it to be $t_s=1$s. 
Since GRBs are currently associated with more massive progenitors we consider $C_1=3,9$.    

To account for rotation of the pre-supernova star the accreting mass is
attributed with specific angular
\begin{equation}
l = l_0 \sin^2\theta
\label{b5}
\end{equation}
where $l_0=10^{16}\mbox{cm}^2\mbox{s}^{-1}$. This is slightly higher then the
equatorial value of $l$ on the PNS surface, $\simeq
7\times10^{15}\mbox{cm}^2\mbox{s}^{-1}$ but still low enough to prevent the development 
of accretion disk around the magnetar. In fact, the rotation does not seem to have 
a strong effect on the solution.

Eqs.(\ref{b1},\ref{b2},\ref{b5}) are also used to set the flow variables in the
ghost cells of the outer boundary.

\subsection{Radiation bubble}

The radiation bubble is assumed to extend up to $r_b=300$~km or 
$r_b=200$~km in different models. The radial and polar velocity components of matter in
the bubble are set to zero and the azimuthal component is calculated using
eq.(\ref{b5}). The prescribed density and pressure distributions are
\begin{equation}
\rho = \rho_b \left(\frac{r}{r_0}\right)^{-a},
\label{c1}
\end{equation}    

\begin{equation}
p = \frac{GM\rho_b}{r_0} \left(\frac{r}{r_0}\right)^{-(1+a)} \left[
\frac{1}{1+a} + \frac{2A}{2+a} \frac{GM}{rc^2} \right] + p_b.
\label{c2}
\end{equation}
For $A=1$ these distributions correspond to a hydrostatic equilibrium.  
In the simulations we use $\rho_b=0.08\rho_0$ and vary $p_b$ to get the desired amount
of thermal energy stored in the bubble, $E_b=C_2 10^{51}$erg. The parameters of four 
models described in this paper are summarized in Table 1.

\begin{table}
   \begin{tabular}{|c|l|l|l|l|}
   \hline
    Model & A & B & C & D \\
   \hline
    $C_1$ & 3 &  3 & 9 & 9 \\
    $C_2$ & 1 & 0.1 & 1 & 0.1 \\
   \hline
   \end{tabular}
\caption{Parameters of the models used for test simulations.}
\end{table}

\subsection{Microphysics}

We use realistic equation of state (EOS) that takes into account the
contributions from radiation, lepton gas including pair plasma, and
non-degenerate nuclei - we have incorporated the EOS code HELM kindly provided
by Frank Timmes for free download
(http://www.cococubed.com/code\_pages/eos.shtml). In this code the contribution
from the electron-positron plasma is computed via table interpolation of the
Helmholtz free energy \cite{TS00}.  The neutrino transport is treated in the
optically thin regime. The neutrino cooling is computed using the interpolation
formulas given in Ivanova et al.\cite{IIN69}, for the Urca-process, and in
Schinder at al.\shortcite{S87}, for pair annihilation , photo-production, and
plasma emission. The neutrino heating rates are computed using the prescriptions
of Thompson et al.\shortcite{TBM01} which take into account gravitational
redshift and geodesic bending. We ignore the Doppler effect due to plasma motion
as its speed relative to the grid never becomes highly relativistic (see
Sec.\ref{results}).  Photo-disintegration of nuclei is included via modifying
EOS (e.g. Ardeljan et al.2005). The equation for mass fraction of free nucleons
is adopted from Woosley \& Baron~\shortcite{WB92}.

\section{Results}
\label{results}

Figure~\ref{f1} shows the global structure of model A at time $t\simeq 200$~ms
when the simulations were terminated because the shock wave has reached the
outer boundary of the computational domain. The low density ``column'' along the
symmetry axis reveals the volume occupied by the collimated outflows from
the magnetar. One can see that these jets have already ``drilled'' holes in the supernova
envelope, that is the layer of stellar material compressed by the supernova shock, 
and are beginning to propagate directly through the collapsing star. 
The unit length in this and other figures is ${\cal L}=GM/c^2 \simeq
2$km.  Thus, the propagation speed of the jets is about
$5\times10^4\mbox{km}\,\mbox{s}^{-1}$ or $0.17c$.  At this speed the jets would
travel across the star of radius $\simeq 2\times 10^5$km in about 4~s.  The
right panel of fig.\ref{f1} shows that the flow speed of the jets can be as high
as $0.5c$. The plots show no signs of the jet termination shock - this is
because the jet flow is sub-fast and can decelerate smoothly when it reaches the
jet head.  The knotty structure of jet is not due to internal conical shocks or
the pinch instability but due to the non-stationary nature of the central
engine. The dead zone of magnetar's magnetosphere opens up in a quasi-periodic
fashion with a characteristic time of 20ms. This causes significant
restructuring of the open magnetosphere and variability of the outflow. This is
also the origin of waves clearly visible in the velocity field within the
circular region of 3000km radius (right panel of fig.\ref{f1}). One can also see
weaker outflow along the equatorial current sheet of magnetars magnetosphere.

\begin{figure}
\includegraphics[width=85mm]{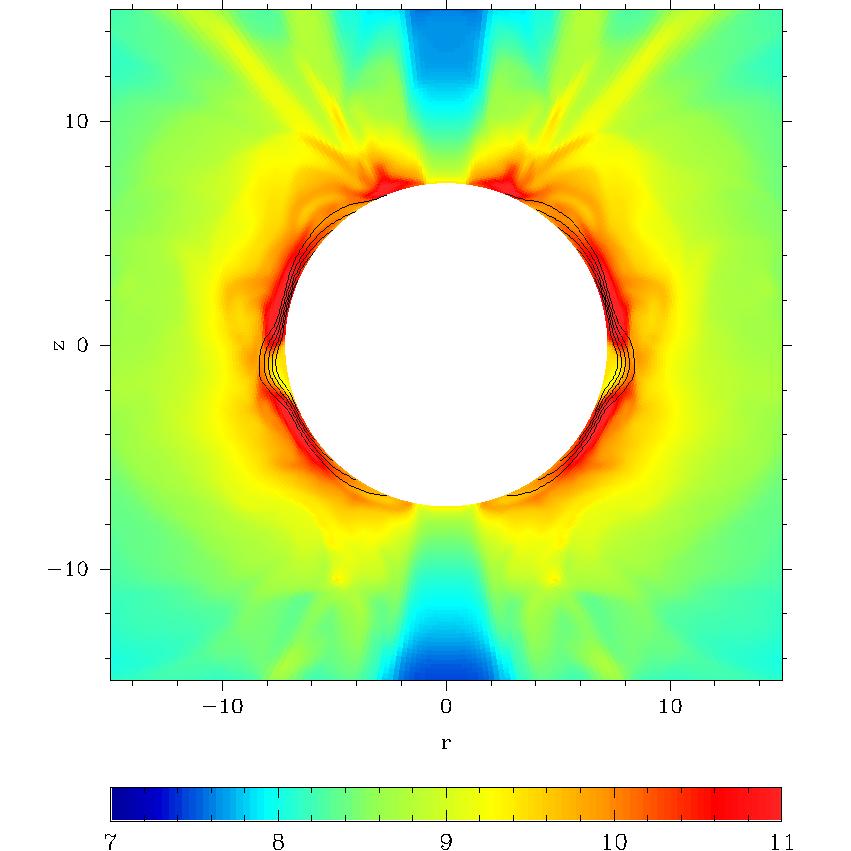}
\caption{Model D at time $t=169$ms. The colour image shows $log_{10}\rho$ and the contours
show the magnetic field lines. 
The unit length in all figures in this paper is ${\cal L}=GM/c^2 \simeq 2$km.
}
\label{f2b}
\end{figure}
\begin{figure}
\includegraphics[angle=-90,width=80mm]{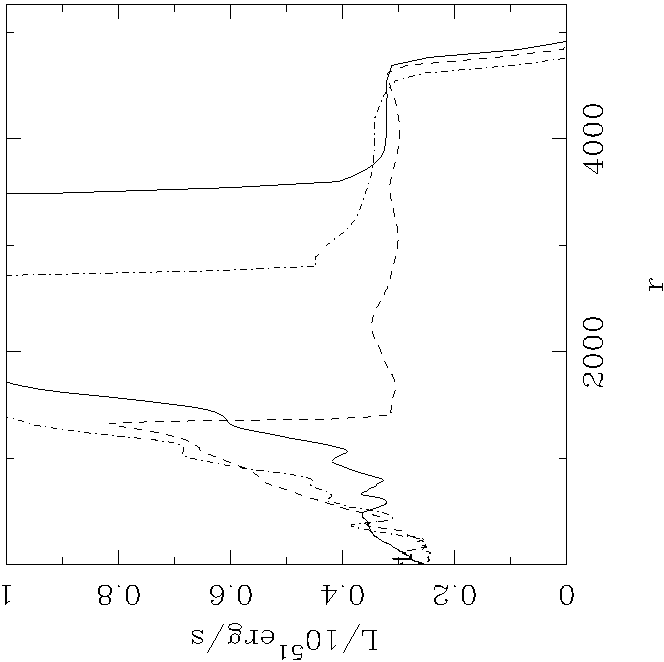}
\caption{Total free energy flux through sphere of radius $r$ for model A (solid
line), model B (dashed line), and model C (dash-dotted line).}
\label{f3}
\end{figure}

Figure~\ref{f2} shows the global structure of model B at the same time.  The
most apparent difference is in the size of the quasi-spherical supernova
shock. It propagates noticeable slower because of the lower energy deposited in
the radiation bubble. However, the propagation speed of the jets is almost the
same as in model A suggesting that their power must be similar as well.  One can
also see that inside the supernova envelope the jets propagate inside a relatively
thin channel and outside of the envelope they propagate through a low density
cocoon inflated during the previous evolution. A similar cocoon is beginning to
form in model A (see fig.\ref{f1}) suggesting that at later times models A and B
could look even more alike.

Figure~\ref{f2a} shows the global structure of model C at time $t=277$ms. Since the ram
pressure of the infalling material is higher than in models A and B the jet propagation
speed is somewhat lower. The same factor explains lower propagation speed of the supernova 
shock in this model compared to model A. In other respects the solution is not much 
different.

The evolution of model D starts in very much the same fashion as for other models but at later 
times we observe a substantial fall-back of the shocked stellar material on the magnetar. 
The neutrino-cooled accreting matter eventually blocks the jet channel and squashes the 
magnetosphere against the star surface (see fig.\ref{f2b}). Soon after the code crashes 
as the conditions near the surface become too extreme. 

Figure~\ref{f3} shows the total flux of ``free energy'', the total energy at
infinity minus the rest mass energy, as a function of radius for
models A,B, and C. The ``spikes'' at $r\simeq 1000$ for model B, between $r=1000$ and
$r=4000$ for model A, and between $r=1000$ and $r=3000$ for model C correspond to 
the blast wave of normal supernova explosion. The ``plateaus'' to the
left and to the right of the spikes correspond to the jets. Thus, in all models
the spin-down power released in the jets is indeed more or less the same,
$L\simeq 3\times 10^{50}$erg/s.  The corresponding spin-down time is $\simeq
37$s suggesting that most of the rotational energy will be released after the
jets break out from the progenitor star. While this is a promising result for
the magnetar model of GRBs it also suggests that only a modest fraction of the
total rotational energy of the magnetar can be transferred to the supernova
ejecta.  In all three models the ratio of Poynting flux to the rest mass energy flux
is $\sigma\simeq 2.5$.  This is approximately 5 times higher than the value predicted by the
theory of unconfined equatorial magnetar wind (see eq.(27) in Metzger et
al.\shortcite{MTQ07}).

\begin{figure*}
\includegraphics[width=57mm]{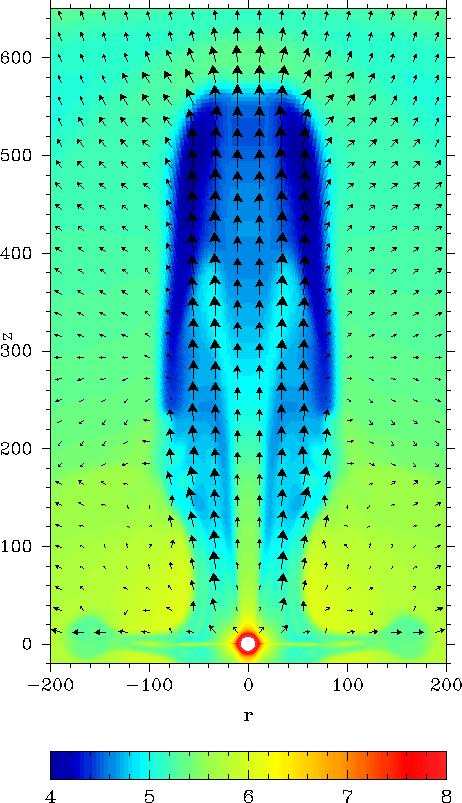}
\includegraphics[width=57mm]{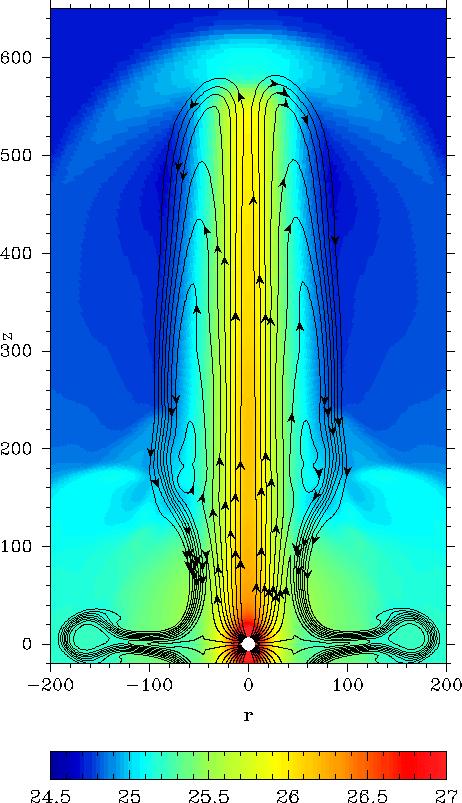}
\includegraphics[width=57mm]{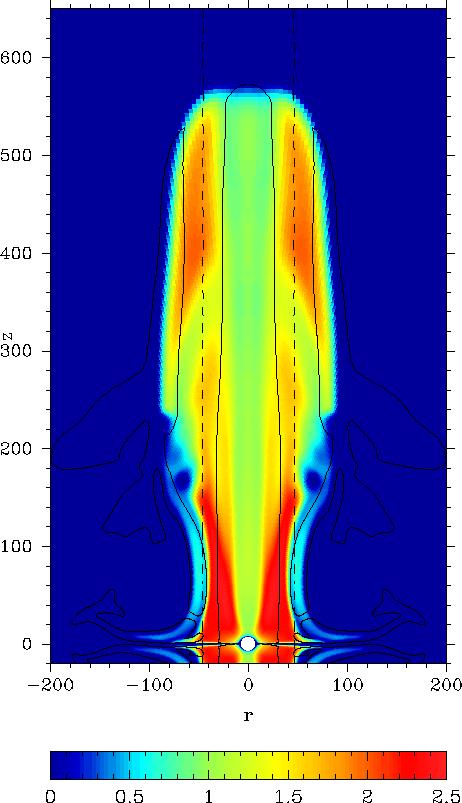}
\caption{Model A at $t\simeq 21$~ms.  {\it Left panel}: $\log_{10}\rho$ and the
poloidal velocity field; {\it Middle panel}: The colour image shows
$\log_{10}p\sub{tot}$, total pressure, and the contours show the magnetic flux
function; {\it Right panel}: The colour image shows
$\log_{10}(p\sub{m}/p\sub{g})$, the ratio of magnetic pressure to the gas
pressure. The solid lines show the surfaces where the radial phase speed of
ingoing Alfv\'en waves is zero - the one closest to the symmetry axis is the
``Alfv\'en horizon''. The dashed line shows the location of the light cylinder.
}
\label{f4}
\end{figure*}

Figure~\ref{f4} shows model A at the early times, $t\simeq 21$~ms. 
One can see that the magnetar is already producing an outflow which has blown 
a cavity in the shape of a column inside the collapsing star. The radius of this 
column is close to the radius of magnetar's light cylinder, 
$R\sub{LC}\simeq 50MG/c^2 = 100$~km,
which is shown in the right panel of fig.\ref{f4} by the dashed line. We note, however,
that for $z<100$ the column is still inside the light cylinder. The total pressure of the 
outflow is dominated by the magnetic field, the ratio of magnetic to gas pressure 
varying between 10 and 300. One can see strong axial compression of the outflow 
due to the hoop stress of azimuthal magnetic field. The distribution of total pressure over 
the boundary of the volume occupied by the outflow has a clear maximum at the point 
of intersection with the polar axis which is obviously related to the axial compression.
This distribution explains the faster expansion of the cavity blown by the outflow in 
the axial direction. In the middle panel of fig.\ref{f4} one can see the bow shock, 
at $z\simeq 600$, driven by the ouflow into the surrounding. There are no indications of 
the termination shock of the outflow itself - this is because the outflow is sub-fast.  
The solid line in the right panel of fig.\ref{f4}
shows the locus of points where the radial component of the phase velocity of 
ingoing Alfv\'en waves becomes positive.  Inside the column these point 
form a closed surface which we call the Alfv\'en horizon of the magnetar-driven 
outflow. Any Alfv\'en wave can propagate from the exterior of this surface  
into its interior. One can see that the Alfv\'en horizon is always located inside 
the light cylinder, which agrees with theory of steady state MHD flows \cite{Oka78}.

\begin{figure*}
\includegraphics[width=57mm]{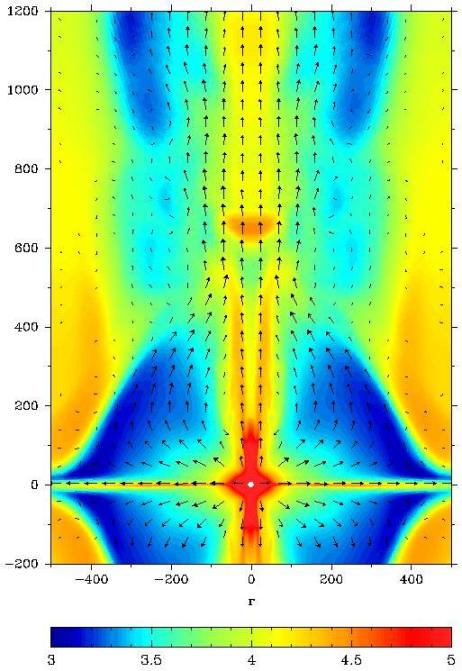}
\includegraphics[width=57mm]{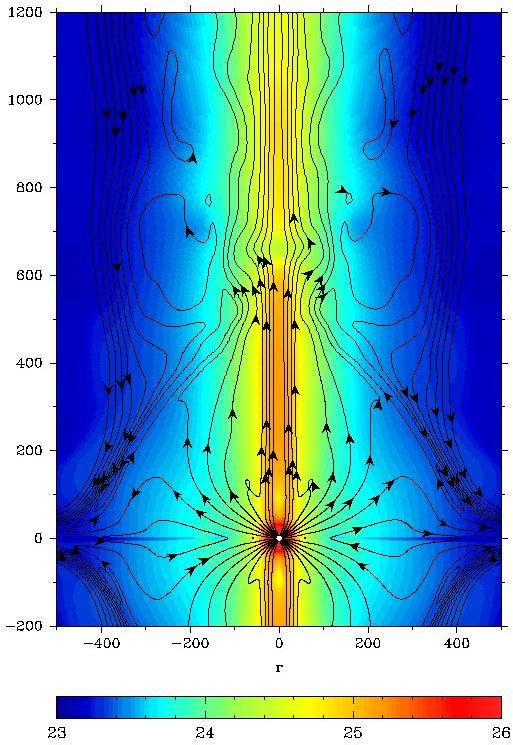}
\includegraphics[width=57mm]{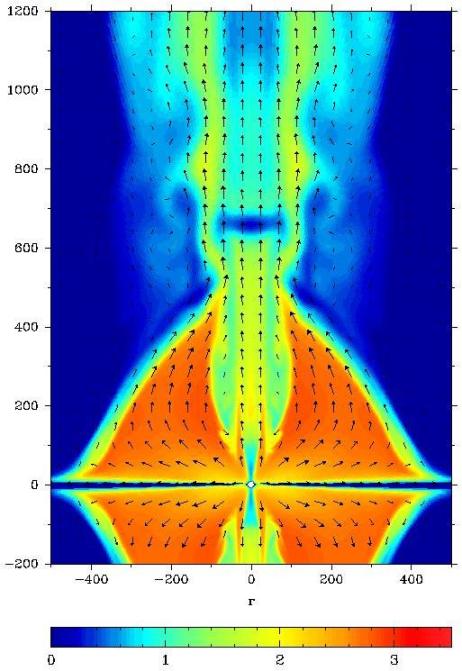}
\caption{Model A at $t\simeq 200$~ms.  {\it Left panel}: $\log_{10}\rho$ and the
poloidal velocity field; {\it Middle panel}: The colour image shows
$\log_{10}p\sub{tot}$, total pressure, and the contours show the magnetic flux
function; {\it Right panel}: $\log_{10}(p\sub{m}/p\sub{g})$, the ratio of
magnetic pressure to the gas pressure and the velocity field.  }
\label{f5}
\end{figure*}

Figure~\ref{f5} shows the inner region of model A by the end of the run
($t\simeq 200$~ms).  By this time the size of the central cavity, that develops
a diamond-like shape, is significantly larger than $R\sub{LC}$, thanks to the
decreased pressure in the surrounding bubble that resulted from its expansion in
the course of supernova evolution. However, the rotational velocity in the
equatorial plane is not ultra-relativistic indicating that the relativistic
winding mechanism is not in operation. The velocity field (left and right panels
of fig.\ref{f5}) shows that close to the magnetar the outflow is similar to an isotropic
wind but then it becomes progressively collimated in the polar direction and
eventually passes through the ``bottle-neck'' nozzles at the axial corners of
the diamond.  The collimation is smooth with no indication of shocks waves which
is expected because the flow is always sub-fast.

\begin{figure*}
\includegraphics[width=50mm]{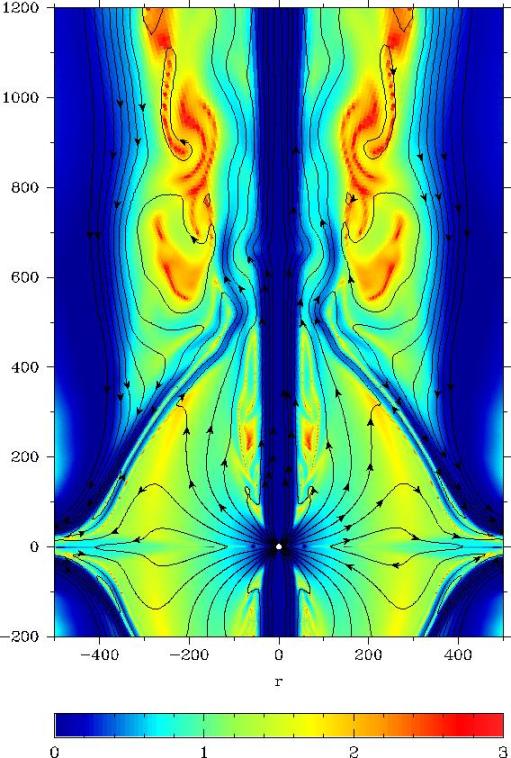}
\includegraphics[width=50mm]{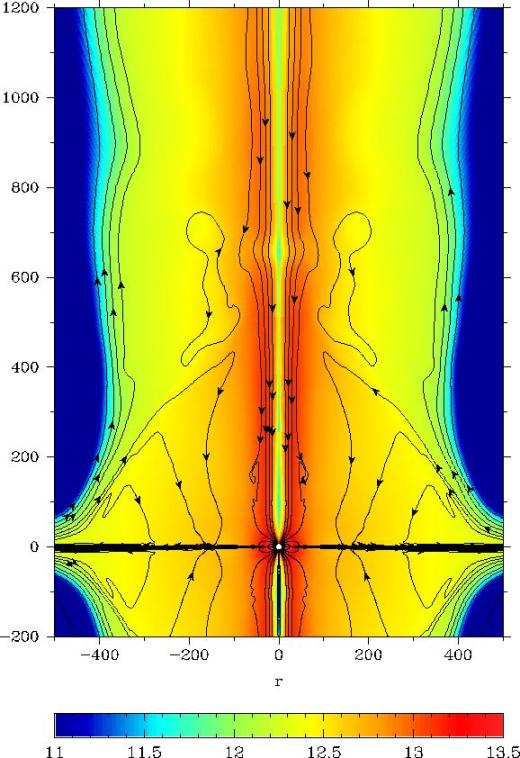}
\includegraphics[width=65mm]{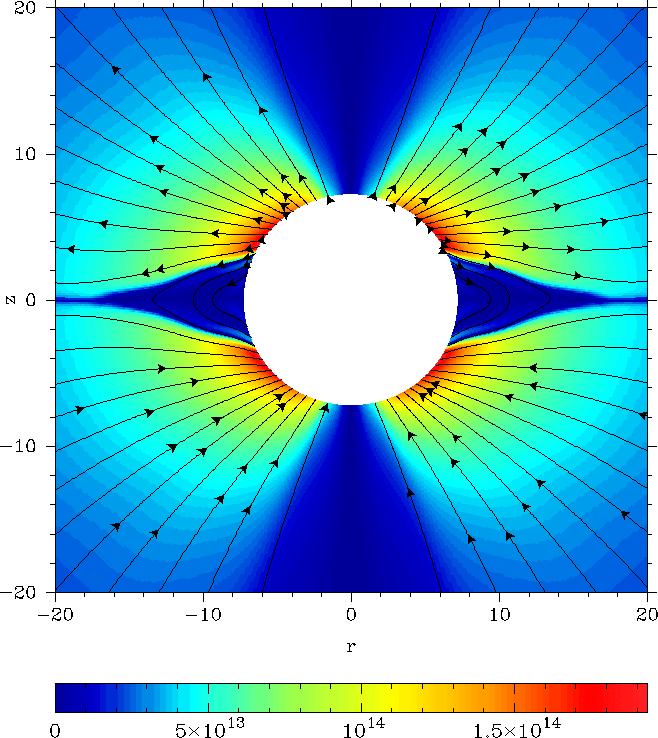}
\caption{Model A at $t\simeq 200$~ms.  
{\it Left panel}: The colour image shows $\log_{10}(|B^\phi|/B_{pol})$, the ratio of azimuthal 
and poloidal components of magnetic field, and the contours show the lines of poloidal magnetic field; 
{\it Middle panel}: The colour image shows the magnitude of the azimuthal magnetic field, 
$\log_{10}|B^\phi|$, and the contours show the lines of electric current. 
{\it Right panel}: The colour image shows the amplitude of $B^\phi$ and the contours show 
the lines of poloidal magnetic field.
The unit length in all figures in this paper is ${\cal L}=GM/c^2 \simeq 2$km. 
}
\label{f6}
\end{figure*}

After passing trough the nozzle each jet enters the channel made during the
previous evolution and propagates inside a low density cocoon. In its outer
layer the cocoon contains poloidal field lines of opposite direction to that of
the jet and closer to the jet one can see magnetic islands and evidence of
vortex motion. These islands are mostly remnants of plasmons injected from the dead
zone of magnetar's magnetosphere (see the right panel of fig.\ref{f6})  
during its oscillatory cycle. In the middle panel of
fig.\ref{f5} one can see one of such recently ejected plasmons.  Close
inspection of animated pictures suggests the following origin of the
oscillations. When the dead zone includes the highest magnetic flux it gradually
accumulates plasma supplied from the magnetar surface. This leads to increase
of the centrifugal force and expanding of the dead zone. Its magnetic field
lines stretch in the equatorial direction trying to open up. The process seems
to accelerate as the dead zone plasma moves further out from the star and
finally escapes in eruptive manner leaving behind thin equatorial current
sheet. Then the oppositely directed field lines of the sheet reconnect and this
restores the initial configuration of the dead zone (a similar behavior of PNS
magnetosphere was reported in Bucciantini et al., 2006).  The amplitude of these
oscillations is large so one gets the impression of the magnetosphere
``breathing heavily''. Since here we integrate the equations of {\it ideal} RMHD
it is the numerical resistivity that is responsible for the observed
reconnection and the reliability of its time-scale is unclear.

The plots of fig.\ref{f5} also reveal a rather peculiar time-dependent structure
inside the diamond cavity and near to the polar axis, to which we will refer as
the ``trap zone''.  It begins at $z\le50$ and continues almost up to the top
corner of the cavity. In the right panel of fig.\ref{f5}, that shows the ration
of magnetic to gas pressure (the inverse of the traditional magnetization
parameter $\beta$), it looks as the region of lower magnetization.  The density
and velocity plots of fig.\ref{f5} show that the trap zone has a relatively high
density and low speed. In fact, the radial velocity in the trap zone changes
sign both in space and time - the density clumps seen in the trap zone exhibit
oscillatory motion along the polar axis.  This explains the peculiar turns of
magnetic field lines in the trap zone (see the middle panel of fig.\ref{f5}).
From time to time plasma clouds are ejected from this zone and the time-scale of
this ejections is similar to the time-scale of global magnetospheric
oscillations.  One of such clouds is seen in density plot of fig.\ref{f5} at
$z\simeq 650$. Another peculiarity of this zone, to which we have no
explanation, is that it is separated from the polar axis by cylindrical shell,
of radius $R\simeq R\sub{LC}$, with a relatively high velocity.

Figure \ref{f6} provides with additional information on the magnetic structure of 
this solution.  One can see that the azimuthal magnetic field dominates almost 
everywhere in the ``diamond zone'' and in the jet, with the exception of the region 
immediately around the symmetry axis. $B^\phi/B_p$ is particularly high in the middle 
of the magnetic cocoon where the poloidal field changes direction (see the left 
panel of fig.\ref{f6}). However, the azimuthal field changes sign only in the equatorial
current sheet. This is clearly seen in the middle panel of fig.\ref{f6} which shows 
that $B^\phi$ does not vanish anywhere else. The electric current is concentrated in 
the jet core, the cocoon surface, and the equatorial current sheet thus showing that 
within most of the jet volume the magnetic field is almost force-free.  The right panel 
of fig.\ref{f6} shows the most inner part of the magnetar magnetosphere and reveals 
the existence of its dead zone.        

\section{Discussion}
\label{discussion}

\subsection{Cautionary notes}

The main motivation of this study was to explore the effects that the birth of a
millisecond magnetar may have on the development of delayed supernova
explosions. The problem of supernova explosions is extremely complicated and we
had to make a number of simplifying assumptions driven by the limitations of our
numerical method. The most problematic issue is the initial setup where we
assume that the magnetic field of the magnetar generated via $\alpha$-$\Omega$-dynamo
or other dynamo has just emerged above magnetar's surface and the large amount of
energy needed to drive standard supernova explosion have just been deposited in
the radiation bubble.  Although the time-scales of these processes are similar
($\le 1s$), even small asynchronism can significantly change the initial
conditions simply because of the large expansion speed of the supernova
shock. This has to be kept in mind when analyzing our numerical models,
particularly their early evolution.  However, we expect the models to be more or
less reliable in the second half of the simulation runs.  

Another important
limitation of our models, as well as many other models of magnetically-driven
astrophysical flows, concerns the potentially destructive role of the kink mode
instabilities. Obviously, these instabilities are totally suppressed in our
axisymmetric simulations and full 3D simulations are needed to investigate this
issue. We can only comment on two stabilizing factors present in the current
setting.  Close to the magnetar the poloidal magnetic field may provide strong
``backbone'' support for the outflows \cite{AZKB}. Further away the jets
propagate within a channel with ``high-inertia walls'' which could effectively
dump the kink-type motions inside the channel.  

Finally, the condition of
axisymmetry means that we can only study the exceptional case of aligned
rotator. However, we do not expect the case of oblique rotator to give
dramatically different results. For example, in the limit of Magnetodynamics
(inertia-free Relativistic MHD) the spin-down power of orthogonal rotator
exceeds that of aligned rotator but only by a factor of two \cite{S06}

\subsection{Comments on the mechanism of explosion}

Many features of the magnetar driven outflows observed in our simulations are 
similar to those predicted by Uzdensky \& McFadyen \shortcite{UM06}. 
There is however a number of important differences and one of them is the nature 
of explosion.  None of the succesful explosions in our simulations are 
magnetically driven. These explosions are produced via the deposition of  
thermal energy in the radiation bubble of the initial solution. We have never seen 
the long phase (many rotational periods) of building up of magnetic pressure behind 
the stalled shock -- the key process in their scenario. 
Moreover, in the simulation runs where the 
amount of energy deposited in the buble was reduced in order to approach the 
conditions of failed supernova and to engage the magnetic mechanism the magnetar 
magnetosphere did not expand but was pressed back to the magnetar surface under 
the weight of neutrino-cooled layers of accreting material. It is easy to see 
why. Suppose that mass $\Delta M$ passes through the stalled shock of failed 
supernova and accumulates at the radius of the light cylinder
$r\sub{LC}=100$km (the critical radius in the magnetic mechanism). 
Then its weight will be $ N=GM\Delta M/4\pi
r^4\sub{LC} \simeq 3\times 10^{28}\mbox{g}\,\mbox{cm}^{-1}
\mbox{s}^{-2}$ for the relatively small $\Delta M = 0.01
M_{\sun}$. This is much larger than the pressure of the dipolar
magnetic field at the same radius, $P_m\simeq 1.3\times
10^{26}\mbox{g}\,\mbox{cm}^{-1} \mbox{s}^{-2} $ (for this estimate we
assume that the at the PNS surface $B_0=10^{15}$G).  Similarly, one
finds that the ram pressure of the free-falling stellar material at
the light cylinder is also much larger than the magnetic pressure.
In principle, the magnetic field generated in the PNS can be
transported in the region bounded by the stalled shock of failed supernova by 
the large scale convective motions developing in this region and even to be locally 
amplified via some sort of dynamo. However, in this case the dynamics of magnetic field 
would be much more complicated than the one described in the simple magnetospheric model of 
Uzdensky \& McFadyen~\shortcite{UM06} and in order to capture it one would require much 
more sophysticated simulations than ours.      

\subsection{Magnetised winds and magnetic towers}

Until recently the magnetically driven outflows from astrophysical objects 
were classified only as ``winds'' or ``jets'', depending on whether the outflow 
proceeds in all directions or it is collimated within a cone due external confinement 
of one sort or another. But since the work of Lynden-Bell\shortcite{L96,L03}, who 
proposed a particular magnetic mechanism for generation of astrophysical gets, a new term, 
``magnetic tower'' is gradually gaining popularity too. In many computer 
simulations the development of column-like outflows have been observed and these 
outflows have been pronounced magnetic towers, without a detailed analysis. 
In fact, collimated winds and magnetic towers have 
many similar properties like the magnetic braking of the central rotator, the 
amplification of azimuthal magnetic field, the axial compression due to hoop stress 
etc. A magnetic tower occupies a finite volume which gradually increases in time 
whereas winds are usually associated with steady-state solutions that extend
to infinity. However, once a wind becomes super-fast its interaction
with the external medium does not affect the upstream solution and thus
this distinction is superficial. A magnetic tower is prevented from
lateral expansion by external pressure but some kind of external force
is needed to confine any type of magnetic flow within a finite volume.
The magnetic field lines of magnetic tower have both foot-point
anchored to the rotator but the same can be said about the wind from a
star with, for example, a dipolar magnetic field; the so-called open magnetic
field lines of stellar winds close far in the wind zone. 

From our view the main distinction between a magnetic tower and a 
magnetically driven wind is in the way the azimuthal magnetic field is
generated in these flows. 
In the magnetic tower model the azimuthal field is generated solely due the
differential rotation at its base (originally an accretion disk)
which causes winding of closed magnetic field lines whose foot points rotate
with different angular velocities.  Thus, both these foot points are assumed
to be in communication with each other by means of torsional Alfv\'en waves; the
faster rotating foot-point creates stronger wave and wins in the sense
that its rotation determines the sigh of the generated azimuthal
field. This also implies that the magnetic tower is a quasi-magnetostatic 
configuration - the vertical growth of the tower can be considered as 
a sequence of magnetostatic models parametrized by time.  

In contrast, wind solutions are characterized by the appearance of Alfv\'en 
horizon; once the Alfv\'en waves emitted by the rotator cross this surface they 
can no longer come back. Then the closed field lines of rotator's magnetic field may  
form two groups: (1) those that close within the horizon and (2) those that close 
outside of it. If a magnetic field line belongs to the first group 
then its foot points can communicate by means of Alfv\'en waves. In this case it is 
still important whether the rotation at the base is differential or not.    
If, however, it belongs to the second group then the communication is broken 
and for this reason the field line can cosidered as opened. For example, the magnetosphere of 
a solidly rotating star with dipolar magnetic field has two zones: the 
so-called ``dead zone'' where the magnetospheric plasma corotates with the star and the 
magnetic field is purely poloidal, and the ``wind zone'' where the magnetic field lines are 
``opened'' and the magnetic field has an azimuthal component. The magnitude of this 
component depends on the angular velocity of only one foot point, the upstream one, and 
it is created solely by the Alfv\'en waves emitted from this point. One may say that the winding
of magnetic field in the wind zone is the feedback reaction of plasma on the magnetic field 
that tries to spin it up, it is there due to the finite inertia uploaded on the magnetic field 
lines in the wind. There is also finite inertia in the dead zone but in the absence of 
outflow there is also plenty of time to enforce full corotation.              

Formally, the magnetic tower solutions can be constructed for a rather arbitrary 
distributions of external pressure \cite{L03}. However, for real flows with 
finite Alfv\'en speed the lateral expansion due to declining external pressure 
can be a factor that works in favor of wind solutions.       
Indeed, the non-relativistic Alfv\'en speed $v_a\propto B_p/\sqrt{\rho}$. If $\varpi$ is the 
cylindrical radius of a magnetic surface then to zero-order approximation 
we have $B_p\propto \varpi^{-2}$ and, assuming constant mass loading of 
magnetic field lines,  $\rho\propto\varpi^{-2}$. This gives $v_a\propto\varpi^{-1}$
and if the speed of longitudinal expansion does not decrease as fast as this then at 
some point the outflow will become super-Alfvenic. This would signify a transition 
to the wind regime.  

Relativity further complicates the matter as it puts an upper limit on
the location of Alfv\'en surface - it must reside inside the light
cylinder~\cite{Oka78}. This seems to be related to the fact that in relativistic MHD 
there is inertial mass associated with the magnetic (or rather electromagnetic) field itself. 
In fact, in the limit of Magnetodynamics, that is relativistic MHD with vanishingly small 
particle inertia (e.g. Komissarov 2002, Komissarov et al.2007), 
all of the inertia is due to electromagnetic field and these surfaces coincide --   
this is why the light cylinder plays such an important role in the theory of 
pulsar winds.  In this respect we wish to comment on the ``pulsar-in-cavity'' problem  
of Uzdensky and MacFadyen\shortcite{UM06}. They consider a magnetized neutron star 
placed in the center of an empty cavity inside a collapsing supernova progenitor. 
The radius of this cavity is assumed to exceed the radius of star's light cylinder. 
By making the cavity so large they effectively create conditions that allow 
the development of a pulsar wind with the kind of magnetic winding that is 
specific only to winds (notice that the star rotates as a solid body). 
Rather unfortunately, however, they still call the outflow that may eventually 
develope in this problem a magnetic tower.  

Summarizing, we wish to stress that both in 
analytical and numerical modeling of magnetically driven ouflows 
it is very important to determine whether the Alfven horizon appears or not. When the 
magnetic tower approximation is used this gives a key test of its validity. When the 
results of computer simulations are analysed this helps to understand the physical nature of 
the outflow. 

\subsection{The nature of magnetar-driven jets}

As the thermally driven explosion (caused by the energy deposition in the radiation 
bubble) develops in our simulations a rarefaction is created in the center and the 
magnetar magnetosphere begins to expand, predominantly in the equatorial 
direction. During this initial expansion an Alfv\'en horizon appears inside the 
magnetosphere and a magnetar wind begins to develop which injects azimuthal magnetic 
flux and energy into the surrounding space. Because the expansion is slow 
compared to the fast magnetosonic speed the fast waves quickly establish 
approximate magnetostatic equilibrium inside the ``magnetic cavity'' and the hoop 
stress of the azimuthal magnetic field ensures that this equilibrium is 
characterised by an axial compression. As a result the normal stress on the cavity 
surface develops a maximum at the poles, the cavity begins to expand 
predominantly in the polar direction and the polar jets are beginning to form 
(see fig.\ref{f4}). The jets remain sub-fast up to the very end of the simulations 
and the termination shock never develops. However, the jet total pressure is 
sufficiently high to drive bow shocks into the surrounding. 
In fact the structure of ouflow is quite similar to that anticipated in 
Uzdensky and MacFadyen\shortcite{UM06}. There are however some qualitative 
differences. First of all, the jets develop before the cavity expands beyond
the light cylinder. This is not very suprizing because the wind magnetization is 
not ultrarelativistic and the Alfv\'en surface can be well inside the light cylinder.
Secondly, there is no highly relativistic plasma rotation in the equatorial plane, 
even when the cavity expands well beyond the light cylinder. 

Eventually, as the central cavity becomes
sufficiently large, one would expect the magnetar wind to become
super-fast and the wind termination shock to appear, as it is assumed 
in \cite{B07a,B07b}.  However, additional studies
are required to determine when exactly this will occur.  This issue may be
further complicated by the magnetar cooling which leads to higher
magnetization of the wind and moves its fast surface further away \footnote{
For inertia free wind this surface moves to infinity}. In any case our results 
suggest that by the time of transition to super-fast regime the magnetar-driven 
jets will already be well established and thus the initial setup used in \cite{B07a,B07b}, 
where the central cavity is assumed to be large and spherical, 
may be somewhat unrealistic.

\subsection{Magnetar as a cenral engine of GRB jets}

Our results show that magnetically-driven outflows from millisecond magnetars
can become highly anisotropic jets at the very early stages of supernova
explosions.  Assuming that their propagation speed does not decrease in time, these
jets would traverse the progenitor star of radius $\simeq 2\times 10^{10}$cm in
about 4 seconds. After this the spin-down energy of PNS will be carried away 
by the jets into the surrounding space and the rate of energy transfer to the 
supernova ejecta will drop. The total power of long duration GRB-jets is not well
known but the observations of radio afterglows which are no longer subject of
relativistic beaming suggest that it can be high, $\simeq 5\times10^{51}$erg
\cite{BKF04}, well in agreement with our results.
Since the combined total power of both jets in our simulations 
is $\simeq 3\times10^{50}\mbox{erg}/\mbox{s}$, prior to the break out the energy 
transferred to the supernova ejecta will be only $\le10^{51}\mbox{erg}$, noticeably 
less than what is usually derived for hypernovae.  
Thus, on one hand our results provide the first self-consistent numerical model of 
a central engine capable of producing, on a long-term basis, collimated jets with
sufficient energy to explain GRBs and their afterglows. On the other hand, the relatively
small amount of energy transferred the supernova ejecta could be a problem due to the 
connection between high velocity SNe and long GRBs. Although the connection 
between SNe and GRBs is supported by strong circumstantial evidence, the direct evidence 
is still scarce with only handful of SNe identified with GRBs. It is quite possible that the
current paradigm connecting GRBs with HNe is simply a reflection of the
observational bias towards more powerful events \cite{WB06,DW06}. 
SN~2002LT/GRB~021211 may the case of a GRB produced by a
standard SN Ib/c \cite{DW06}. 

The relatively high propagation speed of jets in our simulations may in part be
attributed to the condition of axisymmetry. The hydro-simulations of 2D slab
jets (e.g. Komissarov \& Falle 2003) and 3D round jets \cite{N96} show
significantly lower propagation speeds of jets without enforced mirror or axial
symmetry as the kink modes of current-driven or pressure-driven instabilities
result in redistribution of the pressure force over larger area at the jet head
(cf. Aloy et al.,1999).  On the other hand, as the jets enter progressively less
dense outer layers of collapsing star their propagation speed may actually
increase. Slower propagation speed would increase the amount of energy
transferred to the supernova ejecta and reduce the energy available to produce
GRB.

Although our jets are not as fast as required by
the observations of GRBs the gradual cooling of PNS would lead to lower mass loading
and hence higher terminal speed of the jets later on.  
Indeed, according to the computations of Pons et al.\shortcite{PRPLM99} the
neutrino luminosity of PNS decreases by a factor of 10 already during the first
10s and then enters the phase of rapid exponential decline, largely independent 
on the details of initial models and EOS.  Metzger et
al.\shortcite{MTQ07} argue that the magnetization $\sigma=100-1000$ can be
reached when the the PNS still has enough rotational energy to power a
GRB. This corresponds to the asymptotic Lorentz factor 
in the range of $\Gamma_\infty=\sigma/2=50-500$ (e.g. Vlahakis~2004). 
However, the magnetic acceleration of relativistic flows is a rather slow process 
and we do not expect the termination speeds to be reached within the stellar 
boundaries. Further studies are needed to clarify the issue of magnetic acceleration. 

Since the cooling time of PNS is comparable with the break out time,
and the time-scale of long duration GRBs, one would expect that that at the time
of break out the mass-loading of the PNS magnetosphere will still be quite high
and thus the flow speed of the jets will still be relatively low, $\Gamma \simeq
1-10$.  Only later the Lorentz factor will gradually increase and eventually
reach the ultra-relativistic values, $\Gamma\simeq 200$, inferred for the GRB
jets. The implications of such a non-uniform jet structure, with Lorentz factor
gradually decreasing with distance from the the star, for both the prompt and
the afterglow emission of GRBs remain to be investigated.  Here we point out
only few obvious points. First of all one may expect some similarities with the
model of structured jet \cite{MRW98,DG01,Pa05,Ro02,WJ03,KG03}, whose Lorentz
factor varies with polar angle, and the model of two-component jet
\cite{MR01,Vl03,Zh03,RR02,Mi06,Gr06,Ji07,Mo06,Wi07}.  Secondly, since lower
Lorentz factor results in weaker Doppler effect one would expect softer emission
from the parts of the jet emitted earlier and located further away from the
star. This could be the origin of the prompt optical and X-ray emission and
early afterglows. Curiously enough, the model suggests that the zero-point of
gamma-ray bursts lags behind the zero points of prompt X-ray and optical bursts.
Finally, while the collision between the slower earlier jet and the ISM or the
progenitor wind would still produce the strong forward shock usually associated
with afterglows, one might expect a secondary strong forward shock where the
faster late jet collides with the slower early jet.  This secondary shock will
propagate faster than the primary one and will eventually catch up with it. Its
emission, which will be harder and beamed more strongly than that of the primary
shock, may lead to distinctive features in the light curves of afterglows.

It is tempting to consider the global oscillation of magnetar's magnetosphere
and the related non-stationary plasma ejection as the origin of ``inner shocks''
invoked in models of prompt GRB emission \cite{P05b}. However, it is important
to check if the reconnection rate is determined mainly by the global dynamics
and not by the resistivity model (which was purely numerical in our
simulations).  Moreover, it remains to be seen if such oscillations can persist
at later times ($t >$ few seconds) when the the mass-loading of the
magnetosphere drops and the inner ``cavity'' significantly increases in
size. Later, when the jet becomes super-fast, internal shocks can be produced
via interaction with inhomogeneous time-dependent cocoon and instabilities.  The
inhomogeneous structure of the slow jet may also lead to variable emission from
the secondary forward shock.

\section{Conclusions}
\label{conclusions}

The results of our study show that when a core collapse of a massive 
star results in a birth of a millisecond proto neutron star with super-strong
magnetic field (proto-magnetar), this could have a spectacular effect on the 
supernova explosion. During the very early stage of the explosion the magnetar
can produce highly collimated jets capable of 
puncturing the collapsing star in a matter of seconds. The spin-down time of 
the proto-magnetar is an order of magnitude longer thus suggesting that a 
large fraction of its rotational energy, $E\simeq 10^{52}$erg, will be carried 
by the jets into the surrounding space. This supports the idea that at least 
some long duration Gamma Ray Bursts can have  millisecond magnetars as their central 
engines.

The magnetic outflow is best described as a sub-fast super-Alfv\'enic collimated wind. 
The super-Alfv\'enic nature of the flow explains the generation of azimuthal 
magnetic field which soon begins to dominate the flow dynamics. The collimation is a combined 
effect of the inertial confinement by the stellar material and the hoop stress of 
the azimuthal field. The outflow is not a magnetic tower.         

It remains to be seen as to how soon the wind in the central cavity becomes 
super-fast and a proto-PWN is formed inside the collapsing star. The magnetic 
acceleration of the jets inside the channels bored through the star and 
outside of the star is also an important subject for future study. 

It is very likely that the GRB jets produced in this fashion first emerge as
only moderately relativistic flows and only later as the magnetar cools they
become ultra-relativistic. The effect of this on both the prompt and the 
early afterglow emission needs further investigation   

A non-magnetic, e.g. the delayed neutrino-driven explosion of power comparable 
to that of normal core-collapse supernovae is needed to turn-on the magnetic 
mechanism. Otherwise the PNS magnetosphere is unable to expand end develop a 
super-Alfv\'enic wind.  The idea that a failed supernova explosion can be revived 
by a millisecond magnetar does not seem to work.

\section*{Acknowledgments}

We thank Dmitri Uzdensky and for helpful discussion of our results 
and the "magnetic tower" model, Arieh Konigl for many useful comments, 
Frank Timmes for help with implementing HELM EOS package, and the 
anonymous referee for constructive suggestions. 
This research was funded by PPARC under the rolling grant
``Theoretical Astrophysics in Leeds''.


\end{document}